\documentclass[floatfix,twocolumn,prl,amsmath]{revtex4-2}
\usepackage{graphicx}
\usepackage{bm}
\usepackage{amssymb}
\usepackage{dcolumn}
\usepackage{subfigure}
\usepackage{threeparttable}
\usepackage{multirow}
\usepackage{mathrsfs}
\DeclareMathAlphabet{\mathscrbf}{OMS}{mdugm}{b}{n}
\usepackage{color}
\begin{document}
\newcommand {\grbf}[1]{\rm{\boldmath $ #1 $}}
\newcommand{\vn}[1]{{\boldsymbol{#1}}}
\newcommand{\vht}[1]{{\boldsymbol{#1}}}
\newcommand{\matn}[1]{{\bf{#1}}}
\newcommand{\matnht}[1]{{\boldsymbol{#1}}}
\newcommand{\bege}{\begin{equation}}
\newcommand{\ee}{\end{equation}}
\newcommand{\bal}{\begin{aligned}}
\newcommand{\defbar}{\overline}
\newcommand{\SM}{\scriptstyle}
\newcommand{\gretke}{G_{\vn{k} }^{\rm R}(\mathcal{E})}
\newcommand{\gret}{G^{\rm R}}
\newcommand{\gadv}{G^{\rm A}}
\newcommand{\gmat}{G^{\rm M}}
\newcommand{\gles}{G^{<}}
\newcommand{\ghat}{\hat{G}}
\newcommand{\sigmahat}{\hat{\Sigma}}
\newcommand{\glesone}{G^{<,{\rm I}}}
\newcommand{\glestwo}{G^{<,{\rm II}}}
\newcommand{\sigmaret}{\Sigma^{\rm R}}
\newcommand{\sigmales}{\Sigma^{<}}
\newcommand{\sigmalesone}{\Sigma^{<,{\rm I}}}
\newcommand{\sigmalestwo}{\Sigma^{<,{\rm II}}}
\newcommand{\sigmalesthree}{\Sigma^{<,{\rm III}}}
\newcommand{\polarivec}{\boldsymbol{\epsilon}}
\newcommand{\sigmaadv}{\Sigma^{A}}
\newcommand{\Bxc}{\Omega}
\newcommand{\mubo}{\mu_{\rm B}^{\phantom{B}}}
\newcommand{\rmd}{{\rm d}}
\newcommand{\rme}{{\rm e}}
\newcommand{\crea}[1]{{c_{#1}^{\dagger}}}
\newcommand{\annihi}[1]{{c_{#1}^{\phantom{\dagger}}}}
\newcommand{\intkspa}{\int\!\!\frac{\rmd^d k}{(2\pi)^d}}
\newcommand{\eal}{\end{aligned}}
\newcommand{\udot}{\overset{.}{u}}
\newcommand{\exponential}[1]{{\exp(#1)}}
\newcommand{\phandot}[1]{\overset{\phantom{.}}{#1}}
\newcommand{\phandag}{\phantom{\dagger}}
\newcommand{\Trace}{\text{Tr}}
\setcounter{secnumdepth}{2}
\title{
Direct and inverse spin-orbit torques in antiferromagnetic and ferromagnetic FeRh/W(001)
}
\author{Frank Freimuth$^{1,2}$}
\email[Corresp.~author:~]{f.freimuth@fz-juelich.de}
\author{Stefan Bl\"ugel$^{1}$}
\author{Yuriy Mokrousov$^{1,2}$}
\affiliation{$^1$Peter Gr\"unberg Institut and Institute for Advanced Simulation,
Forschungszentrum J\"ulich and JARA, 52425 J\"ulich, Germany}
\affiliation{$^2$ Institute of Physics, Johannes Gutenberg University Mainz, 55099 Mainz, Germany
}
\begin{abstract}
We use \textit{ab-initio} calculations to
investigate spin-orbit torques (SOTs) in FeRh(001)
deposited on W(100). Since FeRh undergoes a
ferromagnetic-antiferromagnetic
phase transition close to room temperature, we
consider both phases of FeRh.  In the antiferromagnetic case we find 
that the effective magnetic field of the
even torque is staggered and therefore ideal to induce magnetization
dynamics or to switch the antiferromagnet (AFM). 
At the
antiferromagnetic resonance
the inverse SOT induces a current density, which can be determined
from the SOT. 
In the ferromagnetic case our calculations predict both even and odd
components of the SOT, which can also be used to describe the
ac and dc currents induced at the ferromagnetic resonance.
For comparison we compute the SOTs in the c($2\times 2$) AFM state of Fe/W(001). 
\end{abstract}

\maketitle
\section{Introduction}
Switching of the N\'eel vector in antiferromagnets (AFMs) by the
spin-orbit torque (SOT) is promising to achieve higher
writing rates on the terahertz scale in magnetic memory
devices (See Refs.~\cite{afmreview_RMP_90_015005,rmp_sot,doi:10.1063/1.4862467} for
recent reviews). Additionally, AFM magnetic memory
allows higher data density than its ferromagnet (FM) counterpart.  
Moreover, it has been proposed that SOT triggers self-sustained
THz oscillations in AFMs~\cite{afm_oscilla_Cheng_Xiao_Brataas_2016}.
N\'eel vector switching through SOT
has been demonstrated experimentally first in the bulk 
antiferromagnets (AFMs) CuMnAs~\cite{ele_switch_afm}
and Mn$_2$Au~\cite{Mn2Au_PhysRevB.99.140409}.
Planar Hall effect and magnetoresistance may be used to read
the state of the AFM bit in the memory device and they have also
been used to obtain an indirect proof of AFM switching. 
Direct evidence of AFM switching is available through
$x$-ray magnetic linear dichroism-photoemission electron
microscopy~\cite{Mn2Au_PhysRevB.99.140409}.

Experimentally, it has been shown that the N\'eel order can be 
switched by the SOT also in bilayers 
composed of a heavy metal layer (HM) and an AFM layer,
such as Pt/Fe$_2$O$_3$~\cite{PhysRevLett.124.027202}.
However, it has been pointed out that
measurements of planar Hall effect and magnetoresistance in
HM/AFM bilayers 
do not always provide clear evidence of
AFM switching~\cite{PhysRevLett.123.247206,absence_evidence_PRL_123_227203}.
Most SOT switching experiments on HM/AFM bilayers considered
insulating AFMs, such as NiO. In CoO/Pt and Fe$_2$O$_3$/Pt bilayers a 
thermomagnetoelastic mechanism rather than the SOT has been
identified as dominating mechanism for current-induced 
switching~\cite{PhysRevLett.123.247206,PhysRevLett.125.077201}.

In this paper we study the SOT in FeRh/W(001) bilayers.
This choice is motivated by three key assets of this system: (i) The use
of W in magnetic bilayers leads to large SOTs~\cite{stt_devices_giant_she_tungsten,ibcsoit}. 
(ii) Epitaxial layers of FeRh(001) can be deposited on W(001) single 
crystals~\cite{FeRh_on_W_Lee_Kao_2010}.
(iii) FeRh has gained considerable interest in the spintronics
community.
Since it exhibits an AFM-FM phase transition close to 
room temperature, femtosecond laser-pulses have been used to
study this phase transition at sub-picosecond
timescales~\cite{doi:10.1063/1.1799244}.
The AFM-FM phase transition in FeRh has also been
used to tune the damping dynamically~\cite{Naneabd2613}.
Lateral spin-pumping between FM and AFM domains
has been found to be crucial for the damping 
enhancement~\cite{spin_pumping_phase_transition_FeRh}.
When FeRh/Pt is excited by femtosecond laser pulses 
superdiffusive spin currents are generated in FeRh and
converted into a charge current by the inverse spin Hall
effect, which leads to a measurable THz 
signal~\cite{seifert_THz_spin_currents_FeRh,medapalli2020femtosecond,2020arXiv200106799L}.
This THz signal varies strongly as FeRh goes through the AFM-FM phase
transition because it is suppressed in the AFM state of FeRh. 
Similarly, the Hall effect changes significantly across the AFM-FM phase
transition~\cite{ahe_ane_ferh}.

Phenomenology suggests that 
the antiferromagnetic resonance is accompanied by
spin-pumping and that conversely 
spin current from the spin Hall effect injected
into an AFM layer exerts staggered effective magnetic fields
on it, which are efficient to induce magnetization
dynamics in the AFM~\cite{PhysRevLett.113.057601,PhysRevB.97.054423,PhysRevB.95.220408}.
However, these concepts have not yet been
investigated by \textit{ab-initio} methods.
Therefore, in this paper we investigate direct and inverse spin-orbit torques in
FeRh/W(001) based on first-principles density-functional theory calculations. 
The direct SOT describes the effective magnetic fields acting on the 
magnetic moments when an electric current is applied, which
may excite magnetization dynamics or switch the AFM or FM
magnetization.
The inverse SOT~\cite{invsot,ac_invsot} describes the electric current induced by
magnetization
dynamics and is therefore related to vertical spin pumping, which has
been
investigated experimentally in the similar system 
of FeRh/Pt~\cite{spin_pumping_phase_transition_FeRh}.
The inverse SOT is also related to the helicity-dependent component of
the THz signal that follows excitation by a fs laser-pulse, and which
was
measured experimentally in FeRh/Pt as 
well~\cite{2020arXiv200106799L,medapalli2020femtosecond}.
For comparison we compute the SOTs also in an Fe monolayer on W(001),
which exhibits a c($2\times 2$) AFM 
configuration~\cite{PhysRevLett.94.087204} 
and is therefore 
compensated like FeRh.

This paper is structured as follows.
In Sec.~\ref{sec_formalism} we
explain first our computational formalism
for the direct SOT (Sec.~\ref{sec_formalism_direct_sot}),
followed by the formalism for the inverse
SOT (Sec.~\ref{sec_formalism_inverse_sot}).
In Sec.~\ref{sec_results} we discuss 
our \textit{ab-initio} results 
in the following order:
In Sec.~\ref{sec_computational_details}
we specify the computational details, and
in Sec.~\ref{sec_even_torkance}
and Sec.~\ref{sec_odd_torkance}
we present our results on the even and odd
direct SOT, respectively, which we obtained in the AFM phase of FeRh/W(001).
We discuss the inverse SOT in the AFM phase of FeRh/W(001) in Sec.~\ref{sec_inverse_sot}.
Next, Sec.~\ref{sec_fm_ferh} is devoted to the direct and inverse
SOT in the FM phase of FeRh/W(001).
Finally, Sec.~\ref{sec_FeW} treats the SOT in Fe/W.
This paper ends with a summary in 
Sec.~\ref{sec_summary}.
\section{Formalism}
\label{sec_formalism}
\subsection{Direct SOT}
\label{sec_formalism_direct_sot}
In order to compute the SOT we use the
formalism described in Ref.~\cite{ibcsoit}.
The torque $\vn{T}_{\mu}$ on atom $\mu$ can be expressed
in terms of the torkance tensor $t_{\alpha\beta,\mu}$ as
\bege
\vn{T}_{\mu}=\sum_{\alpha}\hat{\vn{e}}_{\alpha}
t_{\alpha\beta,\mu}E_{\beta}
\ee
where $E_{\beta}$ is the $\beta$-th component of the applied electric
field
and $\hat{\vn{e}}_{\alpha}$
is a unit vector pointing into the $\alpha$-th Cartesian direction.
We separate the torkance into even and odd parts with respect to
inversion of the magnetization direction.
Since we study antiferromagnets in this work we 
consider the atom-resolved torkances $t_{\alpha\beta,\mu}$,
where the index $\mu$ selects the atom.
The even torkance is given by
\bege
\begin{aligned}
\label{eq_even_torque_constant_gamma}
t^{\rm even}_{\alpha\beta,\mu}=&
\frac{e\hbar}{2\pi\mathcal{N}}
\sum_{\vn{k}n\ne m}
{\rm Im}
\left[
\langle
\psi^{\phantom{R}}_{\vn{k}n}
|
\mathcal{T}_{\alpha,\mu}
|
\psi^{\phantom{R}}_{\vn{k}m}
\rangle
\langle
\psi^{\phantom{R}}_{\vn{k}m}
|
v_{\beta}
|
\psi^{\phantom{R}}_{\vn{k}n}
\rangle
\right]\Biggl\{\\
&\frac{\Gamma
(\mathcal{E}^{\phantom{R}}_{\vn{k}m}
-
\mathcal{E}^{\phantom{R}}_{\vn{k}n})
}{
\left[(\mathcal{E}^{\phantom{R}}_{\rm F}
-
\mathcal{E}^{\phantom{R}}_{\vn{k}n})^2+\Gamma^2\right]
\left[(\mathcal{E}^{\phantom{R}}_{\rm F}
-
\mathcal{E}^{\phantom{R}}_{\vn{k}m})^2+\Gamma^2\right]
}+\\
+&
\frac{
2\Gamma
}
{
\left[
\mathcal{E}^{\phantom{R}}_{\vn{k}n}
-
\mathcal{E}^{\phantom{R}}_{\vn{k}m}
\right]
\left[(\mathcal{E}^{\phantom{R}}_{\rm F}
-
\mathcal{E}^{\phantom{R}}_{\vn{k}m})^2+\Gamma^2\right]
}+\\
+&
\frac{
2
}
{
\left[
\mathcal{E}^{\phantom{R}}_{\vn{k}n}
-
\mathcal{E}^{\phantom{R}}_{\vn{k}m}
\right]^2
}
{\rm Im}\log
\frac{
\mathcal{E}^{\phantom{R}}_{\vn{k}m}
-
\mathcal{E}^{\phantom{R}}_{\rm F}-i\Gamma
}
{
\mathcal{E}^{\phantom{R}}_{\vn{k}n}
-
\mathcal{E}^{\phantom{R}}_{\rm F}-i\Gamma
}\Biggl\}
\end{aligned}
\ee
and the odd torkance is given by
\bege
\label{eq_odd_torque_constant_gamma}
t^{\rm odd}_{\alpha\beta,\mu}=
\frac{e\hbar}{\pi\mathcal{N}}
\sum_{\vn{k}nm}
\frac{\Gamma^2
{\rm Re}
\left[
\langle
\psi^{\phantom{R}}_{\vn{k}n}
|
\mathcal{T}_{\alpha,\mu}
|
\psi^{\phantom{R}}_{\vn{k}m}
\rangle
\langle
\psi^{\phantom{R}}_{\vn{k}m}
|
v_{\beta}
|
\psi^{\phantom{R}}_{\vn{k}n}
\rangle\right]
}{
\left[(\mathcal{E}^{\phantom{R}}_{\rm F}-\mathcal{E}^{\phantom{R}}_{\vn{k}n})^2+\Gamma^2\right]
\left[(\mathcal{E}^{\phantom{R}}_{\rm F}-\mathcal{E}^{\phantom{R}}_{\vn{k}m})^2+\Gamma^2\right]
},
\ee
where $\mathcal{N}$ is the number of $\vn{k}$-points used
to sample the Brillouin zone,
$\mathcal{T}_{\alpha,\mu}$ is the $\alpha$-th cartesian component of the torque operator of atom $\mu$, 
$v_{\beta}$ is the $\beta$-th cartesian component of the velocity operator,
$\Gamma$ is the quasiparticle broadening,
and
$\psi^{\phantom{R}}_{\vn{k}n}$ and $\mathcal{E}^{\phantom{R}}_{\vn{k}n}$
denote the Bloch function
for band $n$ at $k$-point $\vn{k}$ and the
corresponding band energy, respectively.

Experimental works on the SOT typically discuss the
effective magnetic field that one would have to apply
in order to generate a torque of the same size as the SOT.
For a given torque $\vn{T}_{\mu}$ this effective magnetic field
may be computed from
\bege\label{eq_effmagfield}
\vn{B}^{\rm eff}_{\mu}=\frac{\vn{T}_{\mu}\times \hat{\vn{M}}_{\mu}  }{m_{\mu}},
\ee
where $m_{\mu}$ is the magnetic moment of atom $\mu$ 
and $\hat{\vn{M}}_{\mu}$ is its direction.
In order to switch an antiferromagnet, the effective magnetic field
$\vn{B}^{\rm eff}_{\mu}$
needs to be staggered, i.e., its sign needs to be opposite between
antiparallel magnetic moments.

\subsection{Inverse SOT}
\label{sec_formalism_inverse_sot}

While the direct SOT is the generation of a torque
on the magnetization when an electric field is applied,
the inverse SOT is the induction of a current density $\vn{j}$
by magnetization dynamics~\cite{invsot,ac_invsot}.
When this current density
is expressed 
in terms of the atom-resolved torkance,
Eq.~\eqref{eq_even_torque_constant_gamma} 
and  Eq.~\eqref{eq_odd_torque_constant_gamma},
a summation over the atomic site index $\mu$
is required: 
\bege\label{eq_jvol_from_torkance}
j_{\alpha}^{\chi}(t)=\frac{1}{V}
\sum_{\beta,\mu}
t^{\chi}_{\beta\alpha,\mu}(-\hat{\vn{M}}_{\mu}(t))
\left(
\hat{\vn{M}}_{\mu}(t)
\times
\frac{d\hat{\vn{M}}_{\mu}(t)}{dt}
\right)_{\beta},
\ee
where the superscript $\chi$ is set to 'even' and 'odd' to address the
even inverse SOT and the odd inverse SOT, respectively. When the total
inverse SOT is meant, $\chi$ is left blank.

In antiferromagnets with two sublattices we introduce
the two vectors
\bege\label{eq_L_tilde}
\tilde{\vn{L}}(t)=\frac{1}{2}
\left[
\hat{\vn{M}}_{\uparrow}(t)-
\hat{\vn{M}}_{\downarrow}(t)   
\right]
\ee
and
\bege
\tilde{\vn{M}}(t)=\frac{1}{2}
\left[
\hat{\vn{M}}_{\uparrow}(t)+
\hat{\vn{M}}_{\downarrow}(t)   
\right],
\ee
where $\hat{\vn{M}}_{\uparrow}(t)$
and $\hat{\vn{M}}_{\downarrow}(t)$
are the magnetization directions on the two sublattices, which
we denote by $\uparrow$ and $\downarrow$, respectively.
These vectors satisfy $\tilde{\vn{L}}\cdot \tilde{\vn{L}}+\tilde{\vn{M}}\cdot \tilde{\vn{M}}=1$.
Similarly, we define the two torkances
\bege
\tilde{t}^{\chi}_{\beta\alpha}=
\frac{1}{2}
\left[
t^{\chi}_{\beta\alpha,\uparrow}(-\hat{\vn{M}}_{\uparrow})+
t^{\chi}_{\beta\alpha,\downarrow}(-\hat{\vn{M}}_{\downarrow})
\right]
\ee
and
\bege
\bar{t}^{\chi}_{\beta\alpha}=
\frac{1}{2}
\left[
t^{\chi}_{\beta\alpha,\uparrow}(-\hat{\vn{M}}_{\uparrow})-
t^{\chi}_{\beta\alpha,\downarrow}(-\hat{\vn{M}}_{\downarrow})
\right].
\ee
With these definitions the pumped charge current density
in a 2-sublattice AFM can be written
as
\bege\label{eq_pumpedcurr_L_M}
\begin{aligned}
j_{\alpha}^{\chi}(t)&=\frac{2}{V}
\sum_{\beta}
\tilde{t}^{\chi}_{\beta\alpha}
\left(
\tilde{\vn{L}}
\times
\frac{d\tilde{\vn{L}}}{dt}
+
\tilde{\vn{M}}
\times
\frac{d\tilde{\vn{M}}}{dt}
\right)_{\beta}\\
&+\frac{2}{V}
\sum_{\beta}
\bar{t}^{\chi}_{\beta\alpha}
\left(
\tilde{\vn{L}}
\times
\frac{d\tilde{\vn{M}}}{dt}
+
\tilde{\vn{M}}
\times
\frac{d\tilde{\vn{L}}}{dt}
\right)_{\beta}.\\
\end{aligned}
\ee

At the antiferromagnetic resonance the two
magnetization directions $\hat{\vn{M}}_{\uparrow}$
and $\hat{\vn{M}}_{\downarrow}$ precess with slightly
different cone angles, which results in a non-zero
vector $\tilde{\vn{M}}$~\cite{PhysRev.85.329}.
However, usually $|\tilde{\vn{M}}|\ll|\tilde{\vn{L}}|$
is satisfied, which can be used to replace 
Eq.~\eqref{eq_pumpedcurr_L_M}
by approximated expressions.
For a layerwisely compensated layered AFM in an AFM/HM bilayer
with two magnetic sublattices we can approximate
\bege \label{eq_approx_tilde_even}
\tilde{t}^{\rm even}_{\beta\alpha,\ell}=
\frac{1}{2}
\left[
t^{\rm even}_{\beta\alpha,\ell\uparrow}(-\hat{\vn{M}}_{\ell\uparrow})+
t^{\rm even}_{\beta\alpha,\ell\downarrow}(-\hat{\vn{M}}_{\ell\downarrow})
\right]\approx t^{\rm even}_{\beta\alpha,\ell\uparrow}(\tilde{\vn{L}}_{\ell}),
\ee
where $t^{\rm even}_{\beta\alpha,\ell\uparrow}(-\hat{\vn{M}}_{\ell\uparrow})$ 
and $t^{\rm even}_{\beta\alpha,\ell\downarrow}(-\hat{\vn{M}}_{\ell\downarrow})$
are the two torkances of the two sublattices in layer $\ell$ of the
AFM and
\bege
\tilde{\vn{L}}_{\ell}(t)=\frac{1}{2}
\left[
\hat{\vn{M}}_{\ell\uparrow}(t)-
\hat{\vn{M}}_{\ell\downarrow}(t)
\right]   
\ee
is the generalization of Eq.~\eqref{eq_L_tilde} to a layerwisely
compensated layered AFM.
Therefore, in this case the even component of the pumped current
density
can be approximated as
\bege\label{eq_pumpedcurr_even_approx}
\begin{aligned}
j_{\alpha}^{\rm even}(t)&\approx\frac{2}{V}
\sum_{\beta,\ell}
\tilde{t}^{\rm even}_{\beta\alpha,\ell\uparrow}
\left(
\tilde{\vn{L}}
\times
\frac{d\tilde{\vn{L}}}{dt}
\right)_{\beta},\\
\end{aligned}
\ee
where we further approximated $\tilde{\vn{L}}_{\ell}=\tilde{\vn{L}}$.
Similarly, we approximately obtain
\bege\label{eq_pumpedcurr_odd_approx}
j_{\alpha}^{\rm odd}(t)\approx 0
\ee
because 
\bege\label{eq_odd_compensated}
t^{\rm
  odd}_{\beta\alpha,\ell\uparrow}(\hat{\vn{M}}_{\ell\uparrow})=-t^{\rm
  odd}_{\beta\alpha,\ell\downarrow}(\hat{\vn{M}}_{\ell\downarrow})
\ee
when $\hat{\vn{M}}_{\ell\uparrow}=-\hat{\vn{M}}_{\ell\downarrow}$
for a layerwisely compensated layered AFM in an AFM/HM bilayer.

However, this vanishing $j_{\alpha}^{\rm odd}(t)$ is a special case
and 
the odd inverse SOT does not always vanish in AFMs.
Consider for example Mn$_{2}$Au~\cite{Mn2Au_PhysRevB.99.140409,PhysRevB.95.014403} 
and CuMnAs~\cite{ele_switch_afm}.
In these bulk AFMs 
the torkance tensor 
satisfies $t^{\rm
  odd}_{\beta\alpha,\uparrow}(\hat{\vn{M}}_{\uparrow})=t^{\rm odd}_{\beta\alpha,\downarrow}(\hat{\vn{M}}_{\downarrow})$
when $\hat{\vn{M}}_{\uparrow}=-\hat{\vn{M}}_{\downarrow}$.
Thus, we may approximate
\bege
\tilde{t}^{\rm odd}_{\beta\alpha}=
\frac{1}{2}
\left[
t^{\rm odd}_{\beta\alpha,\uparrow}(-\hat{\vn{M}}_{\uparrow})+
t^{\rm odd}_{\beta\alpha,\downarrow}(-\hat{\vn{M}}_{\downarrow})
\right]\approx -t^{\rm odd}_{\beta\alpha,\uparrow}(\tilde{\vn{L}})
\ee
and
\bege\label{eq_pumpedcurr_odd_approx_bulk}
\begin{aligned}
j_{\alpha}^{\rm odd}(t)&\approx-\frac{2}{V}
\sum_{\beta,\ell}
\tilde{t}^{\rm odd}_{\beta\alpha,\uparrow}
\left(
\tilde{\vn{L}}
\times
\frac{d\tilde{\vn{L}}}{dt}
\right)_{\beta}.\\
\end{aligned}
\ee

\section{Results}
\label{sec_results}
\subsection{Computational details}
\label{sec_computational_details}
It has been shown in experiments that
ultrathin epitaxial layers of FeRh(001) can be deposited on W(100) 
single-crystals~\cite{FeRh_on_W_Lee_Kao_2010}.
Since the mismatch between the FeRh (CsCl structure) bulk lattice 
constant of 2.99~\AA\, and the one of bcc W of $a_{\rm W}=3.165$~\AA\, is 5\%,
pseudomorphic growth of the FeRh(001) layer leads to tetragonal
distortion.

\begin{figure}
  \includegraphics[width=\linewidth,trim=11.5cm 24cm 1cm 0cm, clip]{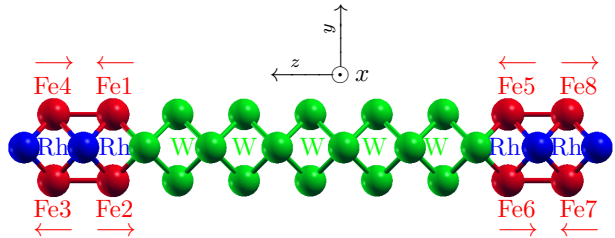}
 \caption{\label{unitcellfigure}
 One unit cell of the FeRh/W/FeRh film. The unit cell is
 repeated periodically along the $x$ and $y$ axes. Red arrows
 illustrate the directions of the magnetic moments on the Fe sites.
}
\end{figure}

We use the film mode~\cite{Krakauer_film_mode} 
in the \textit{ab-initio} program {\tt  FLEUR}~\cite{fleurcode} 
in order
to compute the electronic structure of FeRh on W.
In this mode the unit cell is repeated periodically only in the
in-plane
directions and the resulting film structure is embedded into vacuum.
Since systems with inversion symmetry take less
computational effort in the film mode we consider the 
centrosymmetric system FeRh/W/FeRh, where 11 layers of W(001)
are sandwiched between two layers of FeRh on both sides.
In Fig.~\ref{unitcellfigure}
we show the corresponding unit cell.

In order to compute the antiferromagnetic state we need a magnetic
unit cell with in-plane area twice as large as the one of the crystal
unit cell.
Therefore, we set the in-plane lattice constant to $\sqrt{2}a_{\rm W}=8.459a_{0}$, where $a_{0}$ is
Bohr's radius. In our calculations,
the inter-layer distance is $0.5a_{\rm W}= 2.99a_{0}$ in W
and $2.67a_{0}$ in FeRh. The distance between the W layer and
the Fe layer at the interfaces is 2.83$a_{0}$.
We chose muffin-tin radii of $2.37a_{0}$
in Fe and Rh and of $2.57a_{0}$ in W and performed the calculations
with the generalized gradient approximation~\cite{PerdewBurkeErnzerhof}
 to density-functional theory. Spin-orbit coupling is included in the calculations.
The magnetic moments are 2.63$\mu_{\rm B}$ in Fe1 and Fe2, while they are 
3.21$\mu_{\rm B}$ in Fe3 and Fe4. 

The computational parameters used in our calculations of  a monolayer
of Fe on W(001) (we refer to this system simply by Fe/W(001) in the
following) are given in Ref.~\cite{PhysRevLett.94.087204}. 
Also in this case we compute the inversion symmetric system Fe/W/Fe in
order to reduce the numerical effort.

In order to evaluate the SOT according to
Eq.~\eqref{eq_even_torque_constant_gamma} and
Eq.~\eqref{eq_odd_torque_constant_gamma}
we make use
of Wannier interpolation~\cite{properties_from_wannier_interpolation}
for computational speed-up.
For this purpose we disentangle
18 maximally localized Wannier functions per
transition metal atom, where we employ
our interface~\cite{WannierPaper} between {\tt FLEUR}
and the {\tt Wannier90} program~\cite{wannier90communitycode}.

\subsection{Even torkance}
\label{sec_even_torkance}
\begin{figure}
  \includegraphics[width=\linewidth]{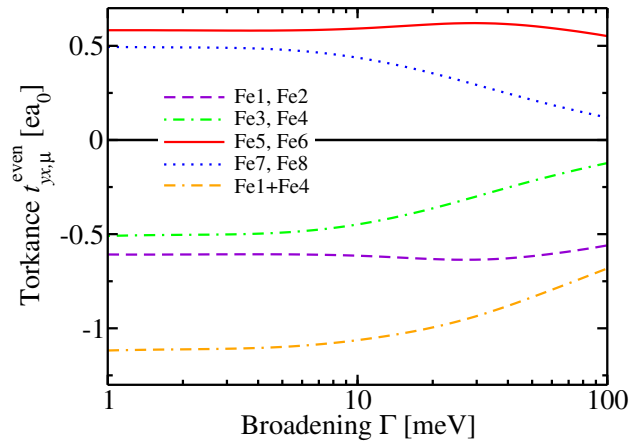}
 \caption{\label{picture_eventorque}
 Atom-resolved even torkances vs.\ quasiparticle broadening $\Gamma$
in the AFM phase of FeRh/W/FeRh.
The atomic site labels Fe1 through Fe8 are explained in
Fig.~\ref{unitcellfigure}.
The sum of the torkances on Fe1 and Fe4 is also shown ('Fe1+Fe4').
The product of elementary positive charge $e$
and Bohr radius $a_{0}$
used as unit of torkance amounts to $ea_{0}=8.478\cdot10^{-30}$~Cm.
}
\end{figure}

We show the atom-resolved even torkance in Fig.~\ref{picture_eventorque}.
While the magnetic moments in Fe1 and Fe2 are aligned
antiferromagnetically,
their torkances agree: $t_{yx,{\rm Fe1}}^{\rm even}=t_{yx,{\rm Fe2}}^{\rm even}$.
Similary, the torkances agree on atoms Fe3 and Fe4, 
i.e., $t_{yx,{\rm Fe3}}^{\rm even}=t_{yx,{\rm Fe4}}^{\rm even}$.
This property of layerwisely compensated layered AFMs is the basis for
Eq.~\eqref{eq_approx_tilde_even}.
Additionally, the four torkances on atoms Fe1 through Fe4 all have the
same sign. Consequently, the effective magnetic field,
Eq.~\eqref{eq_effmagfield} is staggered, i.e., it has opposite sign on 
Fe1 through Fe4 between magnetic moments that point in opposite 
directions. Such a staggered effective magnetic field is precisely
what is necessary to switch the antiferromagnetic layer composed of
Fe1 through Fe4.

Fig.~\ref{picture_eventorque} also shows that the torques on Fe5 and
Fe6 are equal but opposite to the torques on Fe1 and Fe2.
Similarly, the torques on Fe7 and Fe8 are equal but opposite to the
torques
on Fe3 and Fe4. This follows from the fact that 
the space inversion operation maps Fe1 on Fe5,
Fe4 on Fe8, Fe2 on Fe6, and Fe3 on Fe7.
We only show the $yx$-component of
the torkance, because the $xx$ and $yy$ components are zero. The
$xy$-component may be obtained from $t_{xy,\mu}^{\rm even}=-t_{yx,\mu}^{\rm even}$.

In the limit $\Gamma\rightarrow 0$ we find the 
torkances $t_{yx,{\rm Fe1}}^{\rm even}=-0.61ea_{0}$
and $t_{yx,{\rm Fe4}}^{\rm even}=-0.51ea_{0}$.
At high quasiparticle broadening $\Gamma=100$~meV
the torkance on Fe4 is significantly reduced, namely
$t_{yx,{\rm Fe4}}^{\rm even}=-0.12ea_{0}$, while the torkance
on Fe1 is still of similar magnitude, 
namely $t_{yx,{\rm Fe1}}^{\rm even}=-0.56ea_{0}$.
In Ref.~\cite{ibcsoit} we have shown that the even torkance is
described by a scattering-independent mixed Berry curvature in the 
limit $\Gamma\rightarrow 0$. This predicts the even torkance to be
$\Gamma$-independent
at small $\Gamma$.
Indeed at small $\Gamma$, e.g.\ $\Gamma<10$~meV, both torkances are roughly
constant. 

In Ref.~\cite{ibcsoit} we determined the even torkance 
to be $t_{yx}^{\rm even}=-0.83ea_{0}$ in Mn(1)/W(9)
and $t_{yx}^{\rm even}=-0.56ea_{0}$ in Mn(1)/W(15)
in the limit $\Gamma\rightarrow 0$, while at $\Gamma=100$~meV
we found the torkance to be $t_{yx}^{\rm even}=-0.47ea_{0}$
in both Mn/W(001) systems.
Since the in-plane unit cell area of FeRh/W/FeRh is twice as large as the one of Mn/W(001)
we need to compute $t_{yx}^{\rm even, tot}=t_{yx,{\rm Fe1}}^{\rm even}+t_{yx,{\rm Fe4}}^{\rm even}$
in order to perform a meaningful comparison of torkances between
FeRh/W/FeRh
and W/Mn(001), i.e., we have to consider half of the total torque on Fe1
through Fe4. This sum is also shown in Fig.~\ref{picture_eventorque}. 
We find $t_{yx}^{\rm even, tot}=-1.12ea_{0}$ in the
limit $\Gamma\rightarrow 0$ 
and $t_{yx}^{\rm even, tot}=-0.68ea_{0}$ at $\Gamma=100$~meV. 
Thus, the torkances in FeRh/W/FeRh are larger than in W/Mn. However, since
the
sign of the torkances agrees between the two systems, and since the
magnitudes
are similar, we assume that the even SOT is generated by the same
mechanism in both systems. In Ref.~\cite{ibcsoit} we have shown that
the even SOT in W/Mn arises from spin currents and may be attributed
to the spin Hall effect from W. Therefore, we attribute the even
torque in FeRh/W/FeRh also to spin currents from the SHE of W.

The quasiparticle broadening $\Gamma$ also determines the effective
spin-diffusion length. Consequently, we assume that $t_{yx,{\rm Fe4}}^{\rm even}$
decays stronger with increasing $\Gamma$ than $t_{yx,{\rm Fe1}}^{\rm even}$
because the Fe4 is further away from W than Fe1 and therefore a larger
fraction of spin current is lost for high $\Gamma$ before the spin current reaches Fe4.

\subsection{Odd torkance}
\label{sec_odd_torkance}
\begin{figure}
  \includegraphics[width=\linewidth]{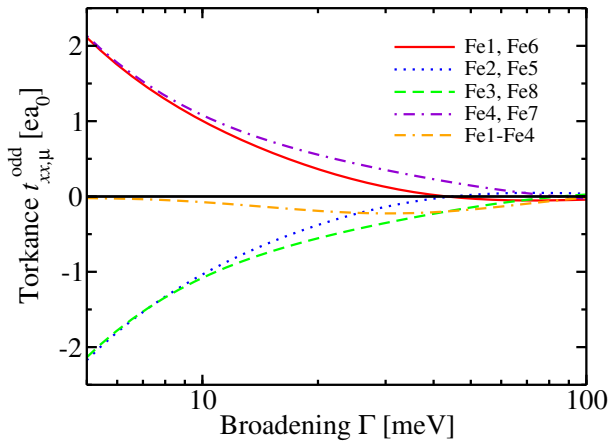}
 \caption{\label{pictureoddtorque}
 Atom-resolved odd torkances vs.\ quasiparticle broadening $\Gamma$ in
 the AFM phase of FeRh/W/FeRh.
The difference of the torkances on Fe1 and Fe4 is also shown
('Fe1-Fe4'), because it can be used to obtain the total odd torkance 
approximately in the FM phase (discussed in Sec.~\ref{sec_fm_ferh}).
}
\end{figure}

In Fig.~\ref{pictureoddtorque}
we show the odd torkance. We only show the component $t_{xx,\mu}^{\rm odd}$, because 
due to symmetry the $xy$ and $yx$ components 
are zero and $t_{yy,\mu}^{\rm odd}=t_{xx,\mu}^{\rm odd}$.
For atoms related by space inversion, e.g.\ Fe1 and Fe5, the torques
are again equal but opposite.
The magnetic moments on Fe1 and Fe2 are antiparallel, but the odd
torkances
at these sites are opposite as well. 
Consequently, the effective field
is
not staggered for atoms Fe1 and Fe2.
Similarly, the magnetic moments
on
Fe3 and Fe4 are antiparallel and their torkances are
opposite such that the effective field is not staggered for these
two atoms either. 
This is a property of layerwisely compensated layered AFMs, which we
expressed also in Eq.~\eqref{eq_odd_compensated}.

On the other hand, Fe1 and Fe4 are antiparallel,
their odd torkances are not staggered, but their effective fields
are staggered. Similarly, Fe2 and Fe3 are antiparallel,
their odd torkances are not staggered, but their effective fields
are staggered.
In order to induce magnetization dynamics in an AFM, the
effective field should ideally be staggered consistently in the AFM.
This criterion is not satisfied by the odd torkance in this system.
The total torkance on Fe1, Fe2, Fe3, and Fe4 is zero for the odd
torque, in contrast to the even torque discussed in the previous section.

At small broadening $\Gamma$ the sign of $t_{xx,{\rm Fe1}}^{\rm odd}$
is different to the one of $t_{xx}^{\rm odd}$ in Mn/W(001), which we
attribute to different interfacial spin-orbit interactions in the two
systems. These differences may be described by opposite signs of
the effective Rashba parameter in the two cases.

\subsection{Inverse SOT}
\label{sec_inverse_sot}
Previous works on spin-pumping in AFM/HM bilayers focused on the dc
spin current pumped at the antiferromagnetic 
resonance~\cite{PhysRevLett.113.057601,PhysRevB.97.054423,PhysRevB.95.220408}.
This dc component is observed when the staggered magnetization is
parallel
to the bilayer interface. Here, we consider a different 
geometry (see Fig.~\ref{unitcellfigure}) with staggered magnetization
perpendicular to the bilayer interface.
In this geometry only ac spin currents can be pumped.
In FM/HM bilayers it has been pointed out that the pumped ac spin current
is larger in magnitude than its dc
counterpart~\cite{ac_voltage_spin_pumping_bauer}.
AC spin currents can even be measured directly~\cite{measure_ac_spin_current}.
Moreover, only the phase-sensitive measurement of the ac inverse SOT
allows us to access both its even and odd
components~\cite{invsot,ac_invsot} in FM/HM bilayers.

Similarly, we expect the ac inverse SOT to induce larger voltages than
its dc counterpart in the present AFM/HM bilayer.  
However, in the present case the odd part
is not easy to access even in a phase-sensitive measurement 
of the ac inverse SOT, because it is approximately zero according to
Eq.~\eqref{eq_pumpedcurr_odd_approx}.
This vanishing odd inverse SOT corresponds to the vanishing total
odd torkance
discussed in the previous Sec.~\ref{sec_odd_torkance}.

Assuming that $\tilde{\vn{L}}$ precesses around the $z$ axis according to
\bege\label{eq_precession_around_z}
\tilde{\vn{L}}(t)
=
\left[
\sin(\theta)\cos(\omega t),\sin(\theta)\sin(\omega t),\cos(\theta)
\right]^{\rm T}
\ee
at the antiferromagnetic resonance,
we obtain from
Eq.~\eqref{eq_pumpedcurr_even_approx}
the current densities
\bege
j_{x}^{\rm even}\approx -\frac{\omega}{V} t_{yx}^{\rm
  even}\sin(2\theta)\sin(\omega t)
\ee
and
\bege
j_{y}^{\rm even}\approx \frac{\omega}{V} t_{yx}^{\rm
  even}\sin(2\theta)\cos(\omega t),
\ee
where 
\bege
t_{yx}^{\rm even}=
t_{yx,{\rm Fe1}}^{\rm even}
+t_{yx,{\rm Fe3}}^{\rm even}=
t_{yx,{\rm Fe1}}^{\rm even}
+t_{yx,{\rm Fe4}}^{\rm even}=
t_{yx}^{\rm even, tot}
\ee
should be used to describe the current induced in FeRh/W(001),
i.e.,  Fe5, Fe6, Fe7, and Fe8 have to be skipped in this summation, because in the
centrosymmetric FeRh/W(001)/FeRh system the inverse SOT
current density is zero if both the upper (Fe1-Fe4) and the lower (Fe5-Fe8) 
AFMs perform the same precession Eq.~\eqref{eq_precession_around_z}.

\subsection{Ferromagnetic FeRh}
\label{sec_fm_ferh}
When we flip all magnetic moments in the system, the even torkance
on a given atom does not change, while the odd torkance on a given
atom changes sign. This holds exactly, because the torkances in
Eq.~\eqref{eq_even_torque_constant_gamma}
and Eq.~\eqref{eq_odd_torque_constant_gamma}
are even and odd, respectively, with respect to inversion of
magnetization, i.e., with respect to flipping all magnetic moments.
When we flip only the magnetic moment of atom $\mu$ but keep
all other magnetic moments unchanged, the even torque on atom
 $\mu$ stays approximately the same, while the odd torque
on atom $\mu$ changes sign and stays approximately the same in
magnitude.
These relations hold only approximately, because by flipping only a
single
magnetic moment we obtain a new system that is not related by any
symmetry operation to the original system.
We may use these approximate relations in order to
describe the ferromagnetic system.
Thus, the even torkances shown in Fig.~\ref{picture_eventorque} for
the
AFM case apply approximately also to the FM case, i.e., the even
torkances
are approximately constant as FeRh passes through the AFM-FM 
phase transition. The torkance may therefore be used to describe the
SOT in the FM case or to compute the voltage induced by the inverse
SOT at the ferromagnetic resonance.

In order to obtain approximately the odd torkances for the FM case, i.e., the
case in which the magnetic moments of Fe2, Fe4, Fe6, and Fe8 are
flipped relative to what is shown in Fig.~\ref{unitcellfigure}, we
only
need to flip the signs of the odd torkances of those atoms.
The resulting total torkance on Fe1 and Fe4 is also
shown in Fig.~\ref{pictureoddtorque} (label 'Fe1-Fe4').
It is the difference between the odd torkance of Fe1 and the one of
Fe4,
because Fig.~\ref{pictureoddtorque} discusses the torkances in the AFM
phase
and therefore we need to multiply the torkance of Fe4 by -1 if we use
it
to describe the FM case. Consequently, the sum of the odd torkances on Fe1 and
Fe4 in the FM phase is approximated by the difference between the odd
torkances
on Fe1 and Fe4 in the AFM phase.

In FeRh/Pt the spin pumping and inverse dc SOT have been investigated
experimentally across the AFM-FM phase transition~\cite{spin_pumping_phase_transition_FeRh}.
Similarly, our results can be used to determine the ac and dc inverse
SOT in the FM phase of FeRh/W.
In the case of FMR-driven magnetization precession around the $z$
axis according to
\bege
\hat{\vn{M}}(t)
=
\left[
\sin(\theta)\cos(\omega t),\sin(\theta)\sin(\omega t),\cos(\theta)
\right]^{\rm T}
\ee
we obtain
the current densities~\cite{invsot}
\bege
j_{x}^{\rm even}\approx -\frac{\omega}{V} 
\left [t_{yx,{\rm Fe1}}^{\rm even}+t_{yx,{\rm Fe4}}^{\rm even}\right]
\sin(2\theta)\sin(\omega t)
\ee
and
\bege
j_{x}^{\rm odd}\approx \frac{2\omega}{V} 
\left[
t_{xx,{\rm Fe1}}^{\rm odd}
-
t_{xx,{\rm Fe4}}^{\rm odd}
\right]
\sin(\theta)\cos(\omega t).
\ee
Here, $\left[
t_{xx,{\rm Fe1}}^{\rm odd}
-
t_{xx,{\rm Fe4}}^{\rm odd}
\right]$
refers to the difference of torkances shown in
Fig.~\ref{pictureoddtorque} (label 'Fe-Fe4').

In Ref.~\cite{2020arXiv200106799L,medapalli2020femtosecond}
a helicity-dependent
THz signal was measured in FeRh/Pt after illumination with a fs
laser-pulse, which can be explained by the model described in 
Ref.~\cite{femtosecond_control_electric_currents_Huisman}.
Similarly,
the odd torkance in the FM phase obtained from our calculations
may also be used to predict a helicity-dependent
THz signal in the similar system FeRh/W. 
For this purpose one may neglect the anisotropy of the odd torkance and
apply our result for magnetization along $z$ to the case with
in-plane magnetization. 

\subsection{Fe/W(001)/Fe}
\label{sec_FeW}
\begin{figure}
  \includegraphics[width=\linewidth]{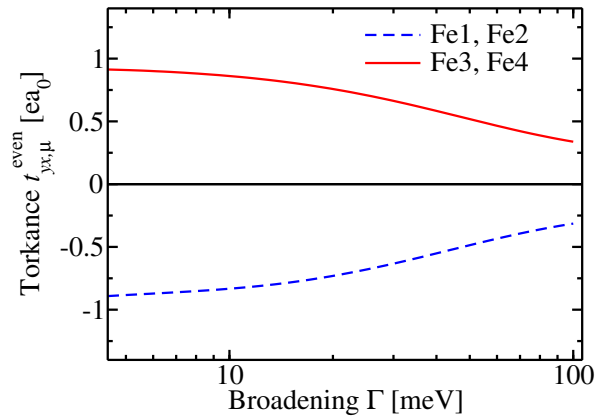}
 \caption{\label{pic_even_FeW}
 Atom-resolved even torkances vs.\ quasiparticle broadening $\Gamma$
in the AFM state of Fe/W(001)/Fe. Fe1 and Fe2 are in the top layer ($z>0$) while Fe3 and Fe4
are in the bottom layer ($z<0$).
}
\end{figure}

In our calculation there are two Fe monolayers in the c(2$\times$2)
AFM state that sandwich the W(001) layer. Each Fe monolayer is
described by an in-plane unit cell containing two Fe atoms, which we
label
as follows:
The top layer ($z>0$) in Fe/W(001)/Fe is composed of the atoms Fe1 and Fe2,
while the bottom layer  ($z<0$) is composed of the atoms Fe3 and Fe4.
In Fig.~\ref{pic_even_FeW} 
we show the even torkance in the AFM
configuration.
At high quasiparticle broadening $\Gamma=100$~meV we find the
torkance $t_{yx,{\rm Fe1}}^{\rm even}=-0.31ea_{0}$, which is slightly
smaller than
the value found in Mn/W(001) of $t_{yx}^{\rm even}=-0.47ea_{0}$~\cite{ibcsoit}.
At $\Gamma=25$~meV we find $t_{yx,{\rm Fe1}}^{\rm even}=-0.68ea_{0}$,
which is larger than $t_{yx}^{\rm even}\approx -0.3ea_{0}$ found in
Fe/Mn(110)
studied in Ref.~\cite{orbisot_dongwook}. However, it is smaller than
$t_{yx}^{\rm even, tot}$ in FeRh/W/FeRh.
The agreement in sign of the even torque between Mn/W(001), Fe/W(001), Fe/W(110) and
FeRh/W(001)/FeRh suggests that it is dominated by the absorption of
spin current from the spin Hall effect of W irrespective of W
orientation (i.e., both in (001) and (110)) and for different FMs and
AFMs, namely for Mn, Fe, and FeRh. However, this behaviour cannot be
generalized to all magnets. For example, it has been shown that in
Ni/W(110) the even torque arises from the orbital torque and that it
is opposite in sign compared to Fe/W(110)~\cite{orbisot_dongwook}.

\begin{figure}
  \includegraphics[width=\linewidth]{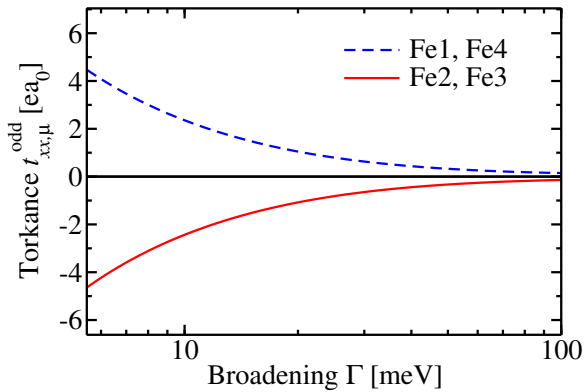}
 \caption{\label{pic_odd_FeW}
 Atom-resolved odd torkances vs.\ quasiparticle broadening $\Gamma$
in the AFM state of Fe/W(001)/Fe. Fe1 and Fe2 are in the top layer ($z>0$) while Fe3 and Fe4
are in the bottom layer ($z<0$).
}
\end{figure}

In Fig.~\ref{pic_odd_FeW}
we show the odd torkance in the AFM
configuration.
At high quasiparticle broadening $\Gamma=100$~meV we find the
torkance $t_{xx,{\rm Fe1}}^{\rm odd}=0.15ea_{0}$, which is opposite in
sign and larger in magnitude when compared with Mn/W,
where we found $t_{xx}^{\rm odd}=-0.082ea_{0}$~\cite{ibcsoit}.
However, the sign agrees with the one of $t_{xx,{\rm Fe1}}^{\rm odd}$
in FeRh/W/FeRh. The sign agrees also with the one of Fe/W(110)
studied in Ref.~\cite{orbisot_dongwook}.

The inverse SOT in Fe/W may be obtained in the same way 
as discussed in Sec.~\ref{sec_inverse_sot}.

\section{Summary}
\label{sec_summary}
We use \textit{ab-initio} calculations in order to study the SOT in 
FeRh/W(001) bilayers. Both the AFM and the FM phase of FeRh
are interesting for spintronics applications. In the AFM phase the
even SOT leads to a staggered effective field, which couples
favourably
to the staggered magnetization. In contrast, the effect of the odd
torque
in the AFM phase is negligible.
We derive expressions for the inverse SOT in AFMs, i.e., formulas
that
express the current-density induced by magnetization dynamics in terms
of the torkance tensors.
We discuss the modifications of the torkance tensor as the system goes
from
the AFM state to the FM state. In the FM phase both even and odd SOT
are significant, and both of them contribute to the ac inverse SOT at
the
ferromagnetic resonance.
For comparison we also compute the SOTs in the c($2\times 2$) AFM 
state of Fe/W(001), where we find smaller even torkances.

\section*{Acknowledgments}
We acknowledge financial support from Leibniz Collaborative Excellence 
project OptiSPIN $-$ Optical Control of Nanoscale Spin Textures, and 
funding  under SPP 2137 ``Skyrmionics" of the DFG. We gratefully
acknowledge 
financial support from the European Research Council (ERC) under the
European 
Union's Horizon 2020 research and innovation program (Grant
No. 856538, project "3D MAGiC”), from the DARPA TEE program through grant MIPR (No.
HR0011831554) from DOI,
and ITN Network COMRAD. The work was also supported by the Deutsche 
Forschungsgemeinschaft (DFG, German Research Foundation) $-$ TRR 173
$-$ 268565370 (project A11), 
TRR 288 – 422213477 (projects B06).  We  also gratefully acknowledge 
the J\"ulich Supercomputing Centre and RWTH Aachen University for
providing 
computational resources under project No. jiff40.

\bibliography{ferh}

\end{document}